# Does DCGAN-Based Synthetic Augmentation Improve Brain Tumor MRI Classification? An Empirical Study

Irhum Jawad Khan
*School of Electrical Engineering and Computer Science (SEECS)*
*National University of Sciences and Technology (NUST)*
*Islamabad, Pakistan*
irhumjawad1@gmail.com

Talha bin Aslam
*Independent Researcher*
*Islamabad, Pakistan*
txlhxv@gmail.com

**Abstract—** Generative adversarial networks (GANs) are increasingly used to augment medical imaging datasets, but synthetic images do not necessarily provide downstream classification benefits. This study investigates whether class-specific Deep Convolutional Generative Adversarial Network (DCGAN) augmentation improves brain tumor classification when the classifier and evaluation set are held constant. Experiments were conducted on 7,200 brain magnetic resonance imaging (MRI) scans across four classes: glioma, meningioma, pituitary tumor, and no tumor. For each class, 1,400 real images were used for training and 400 were reserved for testing. A baseline Swin Transformer classifier was trained using only the real training images and compared with a second model trained using the same real images augmented with 500 DCGAN-generated images per class. Both conditions were evaluated on the identical held-out test set. The two models achieved the same overall accuracy of 96%, while macro F1 remained effectively unchanged and ROC-AUC decreased slightly from 0.987 to 0.982 after augmentation. Class-level analysis showed small redistributions in errors rather than a consistent performance gain. FID values between real and synthetic images ranged from 209.15 to 314.27, indicating substantial distributional differences under the adopted evaluation setup. These results suggest that synthetic augmentation should not be assumed to improve medical image classification and should instead be evaluated for both distributional fidelity and downstream task utility.

## I. INTRODUCTION

Magnetic resonance imaging (MRI) is central to brain tumor assessment, while deep learning has become a common approach for automated tumor classification. Yet medical imaging datasets are often limited by acquisition cost, privacy constraints, and class imbalance, motivating synthetic-data augmentation. DCGANs remain attractive because of their relatively simple training recipe, but reported classification gains do not always isolate the contribution of synthetic images from other changes in training configuration. Recent controlled studies likewise suggest that augmentation benefits can depend on the generator, classifier, and synthetic-to-real ratio [6], [9].

This study asks whether adding DCGAN-generated MRI images improves a fixed downstream classifier when all other experimental factors are held constant. We compare a real-only baseline with a real+DCGAN condition using the same held-out real test set, and independently measure synthetic-image quality using FID. This design separates the effect of data composition from changes in architecture or evaluation protocol.

***Contributions:*** The study (1) provides a matched real-only versus real+DCGAN comparison; (2) reports per-class FID and qualitative real-versus-synthetic inspection; and (3) documents a null result showing no measurable downstream classification gain under the studied conditions. The remainder of the paper reviews related work, describes the methodology, reports results and discussion, and concludes with implications and limitations.

## II. RELATED WORK

GAN-based augmentation has been widely investigated for brain MRI classification. Sandhiya et al. used DCGAN-based reconstruction with segmentation and classification [3], while Haque et al. studied DCGAN augmentation with a Vision Transformer [4]. TumorGANet combined transfer learning and GAN augmentation and reported very high classification performance on a 7,023-image brain-tumor dataset [5]. Alrumiah et al., however, found that the original dataset outperformed DCGAN- and SinGAN-augmented alternatives in their VGG16 experiments [6]. A recent StyleGAN2-ADA study on BRISC 2025 further found that augmentation gains were architecture- and ratio-dependent rather than universal [9]. These findings motivate controlled evaluation rather than assuming that synthetic data are beneficial.

Swin Transformer is a hierarchical vision Transformer using shifted local windows and has been adopted for medical imaging and brain-tumor classification [8], [10], [11]. Here it serves as a fixed downstream model so that the principal experimental variable is training-data composition. For generated-image evaluation, FID measures the distance between real and generated feature distributions [12]. Because the conventional Inception-v3 feature space is derived from natural-image data, FID should not be treated as a direct clinical-quality measure; it is used here as a standardized distributional diagnostic alongside qualitative inspection [14].

## III. PROPOSED METHODOLOGY

### A. Dataset and Sampling

We used the publicly available Brain Tumor MRI Dataset distributed through Kaggle [22], containing four classes: glioma, meningioma, pituitary tumor, and no tumor. The experiment used 7,200 images selected in a balanced protocol: 1,400 real training images and 400 held-out test images per class, yielding 5,600 training and 1,600 test images. The held-out set was kept identical and untouched for both classifier conditions. This explicit sample accounting avoids ambiguity caused by revisions to the hosted dataset over time.

### B. DCGAN Generation

A separate DCGAN was trained for each class at 128×128 resolution. The generator mapped a 100-dimensional latent vector through five transposed-convolution blocks (512→256→128→64→32 channels) with batch normalization and ReLU activations, followed by a Tanh output. The discriminator used five strided-convolution blocks with LeakyReLU activations (negative slope 0.2) and a sigmoid output. Training used Adam with learning rates of $2\times10^{-4}$ for the generator and $1\times10^{-4}$ for the discriminator, $\beta_1$=0.5, batch size 16, and 140 epochs. One-sided label smoothing (real label 0.9), Gaussian instance noise ($\sigma$=0.05, clamped to [−1,1]), and two generator updates per discriminator update were used for stabilization. Weights were initialized with $\mu$=0 and $\sigma$=0.02. After training, 500 synthetic images per class were generated.

### C. Classification Model

The downstream classifier was a Swin Transformer initialized from the public Devarshi/Brain_Tumor_Classification checkpoint [21], which is a fine-tuned Swin-Tiny model. The classification head was reinitialized for the four target classes. Images were resized to 224×224 and normalized using the checkpoint processor statistics. Training augmentation comprised random resized cropping (scale 0.8–1.0), horizontal flipping (p=0.5), rotation (±15°), and mild brightness/contrast jitter (±0.1). The test set received only deterministic resizing and normalization.

### D. Experimental Setup

Two conditions differed only in training-set composition: (A) real-only, with 1,400 real images per class; and (B) real+DCGAN, with the same 1,400 real images plus 500 synthetic images per class. Thus, the augmented condition contained 1,900 images per class. Both models were fine-tuned with the same learning rate ($5\times10^{-6}$), weight decay (0.1), per-device batch size (8), maximum 50 epochs, and early stopping with patience 5 based on validation accuracy. Both runs stopped at epoch 20. Classifier training and FID computation used Kaggle Notebooks with a T4×2 accelerator; DCGAN training used a local CUDA-enabled GPU. Each classifier run required approximately 1–2 h and FID computation across four classes took under 15 min.

### E. Evaluation

Classification was evaluated on the identical 1,600-image real test set using accuracy, per-class precision/recall/F1, macro and weighted F1, confusion matrices, and one-vs-rest ROC-AUC. FID was computed separately for each class between the 1,400 real training images and 500 corresponding synthetic images using an Inception-v3 feature extractor. No synthetic image was used in the held-out test set.

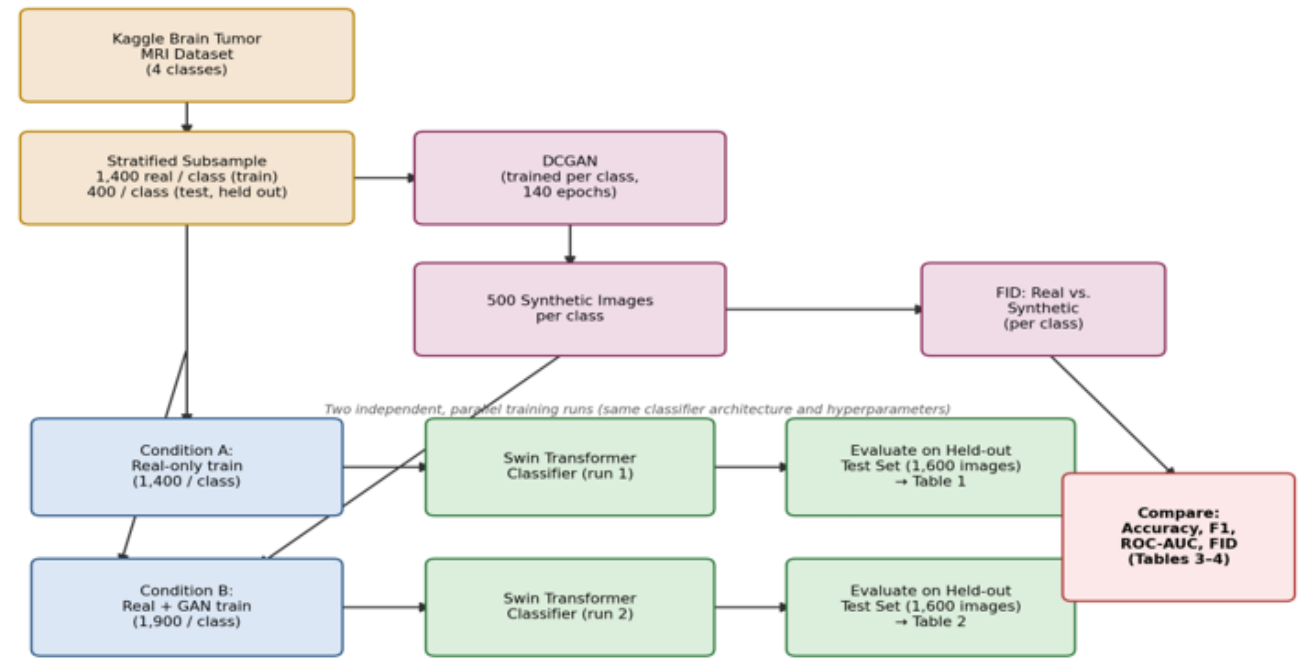

Fig. 1. Controlled experimental pipeline. DCGANs are trained independently by class; the same held-out real test set is used for both classifier conditions, while FID is evaluated independently on real versus generated training images.

## IV. RESULTS AND DISCUSSION

### A. Classification Performance

Table I summarizes the controlled comparison. Both conditions achieved 96% accuracy, macro F1 of 0.96, and weighted F1 of 0.96. ROC-AUC decreased slightly from 0.9954 to 0.9942 after augmentation. At the class level, meningioma F1 increased by 0.01, no-tumor F1 decreased by 0.01, and glioma and pituitary F1 remained unchanged. The confusion matrices in Fig. 2 show a small reduction in glioma→meningioma errors (39 to 31) but a larger increase in glioma→no-tumor errors (18 to 28), leaving the net classification outcome unchanged. The marginal ROC-AUC decrease after augmentation is notable: it suggests that the DCGAN-generated images may have introduced distributional noise that slightly degraded the model's ability to discriminate between classes at varying decision thresholds, even though threshold-specific accuracy remained identical. This is consistent with the high FID values reported in Section IV-B.

TABLE I. CONTROLLED CLASSIFICATION COMPARISON

| Metric | Real-only | Real + DCGAN | Δ |
|---|---|---|---|
| Accuracy | 96% | 96% | 0 |
| Macro F1 | 0.96 | 0.96 | 0 |
| Weighted F1 | 0.96 | 0.96 | 0 |
| ROC-AUC (OvR) | 0.9954 | 0.9942 | −0.0012 |
| Glioma F1 | 0.92 | 0.92 | 0 |
| Meningioma F1 | 0.95 | 0.96 | +0.01 |
| No-tumor F1 | 0.98 | 0.97 | −0.01 |
| Pituitary F1 | 0.99 | 0.99 | 0 |

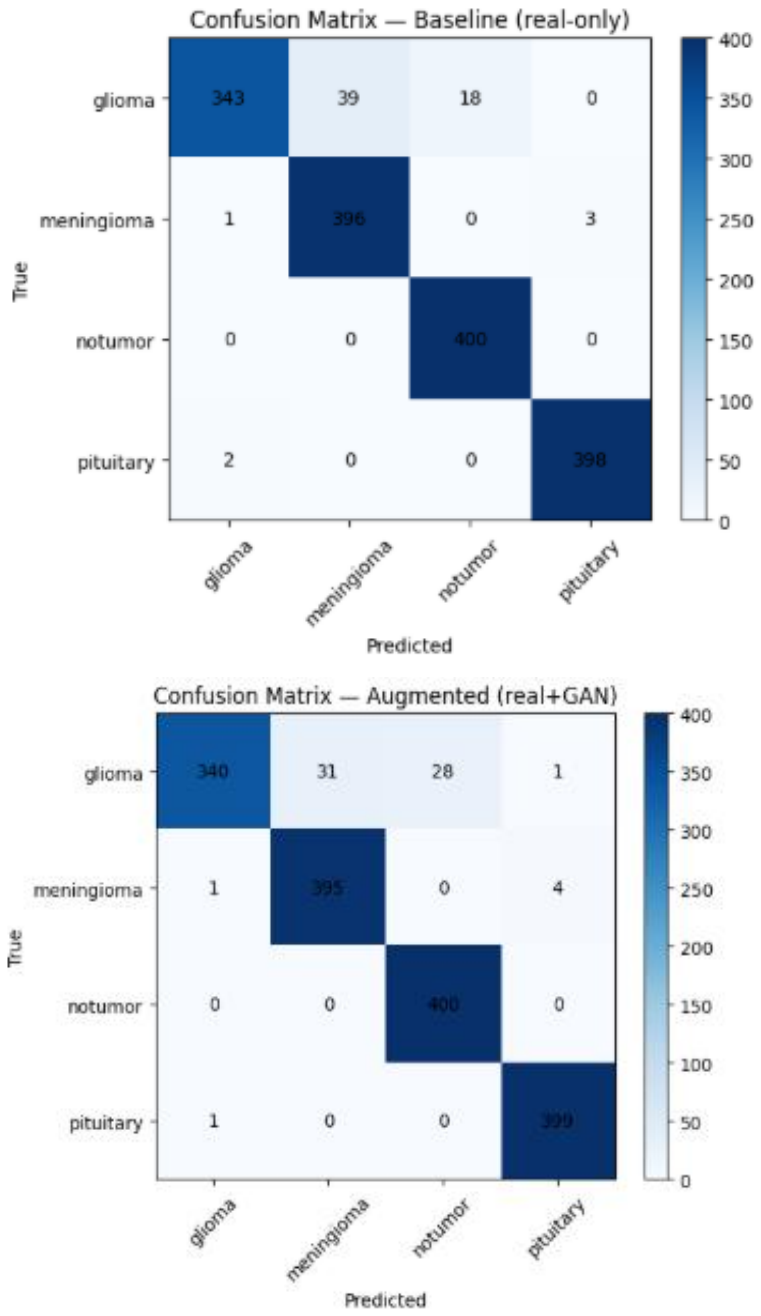

Fig. 2. Confusion matrices on the identical held-out real test set: (a) real-only baseline and (b) real+DCGAN augmentation.

### B. Synthetic Image Quality

The generated-image quality results provide a plausible explanation for the absence of downstream benefit. As shown in Table II, all four FID values are elevated, with the smallest value for no tumor (209.15) and the largest for meningioma (314.27). For reference, FID values below 50 are generally considered indicative of good generative quality on natural image benchmarks; values in the range of 200–315 observed here indicate a substantial real-versus-synthetic distributional gap. Qualitative samples in Fig. 3 show blurred or structurally inconsistent boundaries and reduced internal anatomical coherence in several generated images. These observations are consistent with the distributional gap quantified by FID, although FID alone does not establish clinical realism. The meningioma class produced the highest FID (314.27), which may reflect greater intra-class morphological variability in that category, making it harder for the DCGAN to capture the full distribution. The no-tumor class yielded the lowest FID (209.15), consistent with the relatively uniform appearance of healthy brain tissue compared to tumour-bearing scans.

**TABLE II. PER-CLASS FID BETWEEN REAL AND DCGAN-GENERATED IMAGES**

| Class | FID |
|---|---|
| Glioma | 303.43 |
| Meningioma | 314.27 |
| No tumor | 209.15 |
| Pituitary | 263.85 |

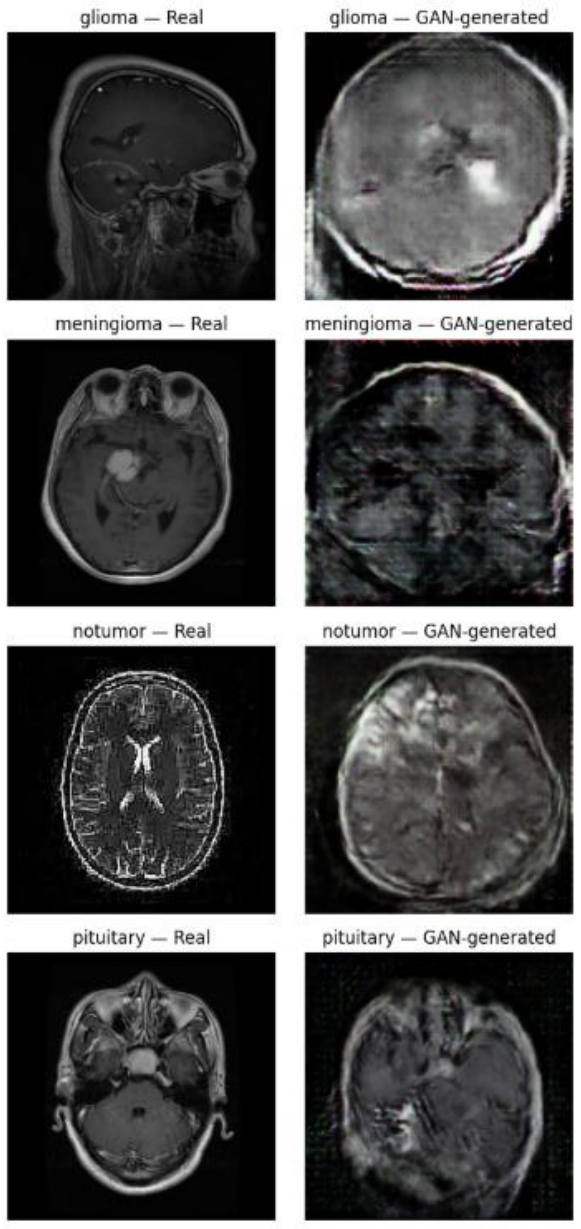

Fig. 3. Representative real (left) and DCGAN-generated (right) MRI images across the four classes. Several synthetic samples show blurred anatomical boundaries and reduced structural consistency.

***Interpretation and limitations:*** The results do not imply that generative augmentation is categorically ineffective. Rather, they indicate that, for this dataset, DCGAN configuration, synthetic-to-real ratio, and Swin Transformer setting, the generated samples did not provide measurable additional discriminative value. Plausible contributors include limited per-class training data, the simplicity of DCGAN relative to newer generators, and the difficulty of reproducing fine MRI anatomy at 128×128 resolution. Important limitations are the use of one public dataset, one generator family, one classifier architecture, one random seed per condition, no formal significance test, no external validation, and no radiologist assessment. Consequently, the 96%-vs.-96% result should be interpreted as a single controlled comparison rather than a statistically general conclusion. Modern generators such as WGAN-GP, StyleGAN2, and diffusion models, together with multi-seed evaluation and external datasets, are natural next steps [15]–[20].

### C. Error Analysis

Inspection of the confusion matrices in Fig. 2 shows that glioma is the most challenging class in both conditions. In the real-only baseline, 39 glioma samples were misclassified as meningioma and 18 as no-tumor, yielding an F1 score of 0.92. After real+DCGAN augmentation, glioma-to-meningioma errors decreased from 39 to 31, while glioma-to-no-tumor errors increased from 18 to 28, leaving the overall glioma F1 unchanged at 0.92. This redistribution suggests that the synthetic glioma samples may have altered the learned class boundaries, reducing confusion with meningioma while increasing overlap with the no-tumor class. The observed pattern is plausible given the visual similarities between glioma and meningioma in some MRI characteristics and the structural inconsistencies observed in the generated samples, as reflected by the glioma FID of 303.43. However, the single-seed experiment does not establish a causal mechanism for this error redistribution.

Meningioma was the only class to show a marginal F1 improvement, increasing from 0.95 to 0.96, whereas no-tumor F1 decreased from 0.98 to 0.97. Pituitary tumor remained the strongest class in both conditions, with an F1 score of 0.99. Although meningioma had the highest FID (314.27), its classification performance improved slightly, illustrating that FID and downstream task performance do not necessarily change monotonically. Overall, the results indicate that DCGAN augmentation redistributed classification errors rather than reducing them, producing no net improvement in overall accuracy or macro-level performance. This finding supports the need to evaluate synthetic medical images not only for distributional similarity but also for their downstream task utility.

### D. Comparison with Prior Work

The null result observed here is consistent with Alrumiah et al. [6], who similarly found that augmenting with GAN-generated brain MRI images did not outperform training on real data alone in a VGG16 experiment. It diverges from Haque et al. [4] and TumorGANet [5], both of which reported classification gains from GAN augmentation; however, neither study employed a matched held-out real test set fixed across augmented and non-augmented conditions, making direct comparison difficult. The StyleGAN2-ADA study on BRISC 2025 [9] likewise found that augmentation benefits were architecture- and ratio-dependent, reinforcing the view that positive results in the literature may reflect experimental design choices rather than a universal property of GAN augmentation. The present study contributes a controlled design that isolates data-composition effects, which is a methodological complement to the performance-oriented framing common in prior work.

### E. Practical Implications

The findings carry practical implications for researchers applying GAN augmentation to medical imaging pipelines. First, FID should be computed and reported as a diagnostic before synthetic images are incorporated into training, since high FID values indicate a distributional mismatch that may limit downstream benefit regardless of visual plausibility. Second, the augmented and non-augmented conditions should use identical held-out test sets to avoid confounding data composition effects with evaluation protocol differences. Third, the choice of generator architecture matters: DCGAN at 128x128 resolution may be insufficient to reproduce the fine anatomical texture of clinical MRI, and newer architectures such as WGAN-GP, StyleGAN2, and diffusion models [15]-[19] offer stronger generation fidelity and warrant evaluation under similarly controlled conditions. These recommendations are intended to support reproducible and rigorous synthetic-data augmentation practice in the medical imaging community.

## V. CONCLUSION

We presented a controlled study of DCGAN-based synthetic augmentation for four-class brain tumor MRI classification. With the classifier architecture, hyperparameters, and held-out real test set fixed, adding 500 DCGAN-generated images per class did not improve accuracy, macro F1, or weighted F1, and slightly reduced ROC-AUC. Per-class FID values of 209.15–314.27 and qualitative inspection indicated a substantial mismatch between generated and real MRI distributions. The central implication is methodological: synthetic medical images should be evaluated for distributional and task relevance before their downstream utility is assumed. Future work should compare stronger generators, multiple seeds, and external clinical datasets.